\documentclass[aps,prl,reprint,groupedaddress,a4paper]{revtex4-1}

\usepackage[latin1]{inputenc}
\usepackage{graphicx}
\usepackage{textcomp}
\usepackage{hyperref}
\usepackage[usenames]{color}
\usepackage{soul}

\newcommand{\fig}[1]{Fig.\,\ref{#1}}
\newcommand{\eq}[1]{equation~(\ref{#1})}

\begin{document}

\title{Microwave cavity-enhanced transduction for plug and play nanomechanics at room temperature}

\author{T. Faust}
\author{P. Krenn}
\author{S. Manus}
\author{J.\,P. Kotthaus}
\author{E.\,M. Weig}
\email[]{weig@lmu.de}
\affiliation{Center for NanoScience (CeNS) and Fakultät für Physik, Ludwig-Maximilians-Universität, Geschwister-Scholl-Platz 1,
München 80539, Germany}

\date{\today}

\begin{abstract}
Nanomechanical resonators with increasingly high quality factors are enabled following recent insights into energy storage and loss mechanisms in nanoelectromechanical systems (NEMS).
Consequently, efficient, non-dissipative transduction schemes are required to avoid the dominating influence of coupling losses.
We present an integrated NEMS transducer based on a microwave cavity dielectrically coupled to an array of doubly-clamped pre-stressed silicon nitride beam resonators.
This cavity-enhanced detection scheme allows resolving the resonators' Brownian motion at room temperature while preserving their high mechanical quality factor of 290,000 at 6.6\,MHz.
Furthermore, our approach constitutes an "opto"mechanical system in which backaction effects of the microwave field are employed to alter the effective damping of the resonators.
In particular, cavity-pumped self-oscillation yields a linewidth of only $5$\,Hz.
Thereby, an adjustement-free, all-integrated and self-driven nanoelectromechanical resonator array interfaced by just two microwave connectors is realised, potentially useful for applications in sensing and signal processing.
\end{abstract}

\maketitle

The increasing importance of nanomechanical resonators for both fundamental experiments\,\cite{OConnell2010,Teufel2011,Chan2011} and sensing applications\,\cite{doi:10.1021/nl801982v,JensenK.:533} in recent years is a direct consequence of their high resonance frequencies as well as low masses. However, because of their small size, they couple only weakly to their environment which can make it difficult to efficiently transduce their motion. This coupling can be strongly enhanced via an optical\,\cite{Metzger2004,bib:Gigan2006,bib:Arcizet2006a,Kleckner2006,Schliesser2008,Eichenfield2009a} or electrical microwave\,\cite{PhysRevLett.99.137205,Regal2008,sillanp:011909,doi:10.1021/nl102771p,Rocheleau2010,hao:113501} cavity.
While both methods enable sensitive displacement detection, only the latter is suitable for large scale integration of many resonators with a single cavity.
Up to now nanoelectromechanical transduction via microwave cavities is predominantly performed at cryogenic temperatures to benefit from superconducting cavities capacitively coupled to superconducting mechanical resonators.
With a main focus on quantum mechanical ground state cooling\,\cite{Teufel2011}, the potential of cavity nanoelectromechanical systems for integrated transduction at room temperature is yet to be exploited. 

We present an approach based on a copper microstrip cavity operating at $300$\,K.
While previous works\,\cite{PhysRevLett.99.137205,Regal2008,sillanp:011909,Rocheleau2010,doi:10.1021/nl102771p} relied on capacitive coupling between cavity and metallised resonator, we employ a dielectric resonator made of highly stressed silicon nitride.
This avoids additional damping by losses in the metallisation layer, which frequently is one of the dominating sources of dissipation at room temperature\,\cite{Sekaric2002,2011arXiv1111.1703Y}.
For transduction, we take advantage of dielectric gradient forces, which are becoming more and more established as a powerful tool to control NEMS\,\cite{schmid:163506,Unterreithmeier2009,Anetsberger2009,PhysRevA.82.023825}:
If a dielectric beam is placed in between two vertically offset electrodes, its vibration will induce a periodic modulation of their mutual capacitance.
We demonstrate that this modulation alters the response of a connected microwave cavity which can be demodulated to probe the displacement of the nanomechanical resonator.
The resulting heterodyne cavity-enhanced detection scheme allows to probe the resonator's Brownian motion with a sensitivity of presently $4.4$\,pm$/\sqrt{\rm Hz}$ at $300$\,K.
We have tested the scheme to operate at temperatures between $300$\,K and low temperatures ($4$\,K), at which superconducting cavities become superior.
Furthermore, the coupled cavity-resonator device is a microwave analogy of an optomechanical system: 
Cavity electromechanics can be employed to amplify or damp the mechanical vibration utilising the dynamical backaction of the microwave field.
By strongly amplifying the motion, the regime of cavity-pumped self-oscillation is reached.
The resulting high-amplitude, narrow-band signal with a linewidth of only $5$\,Hz can be used to track the resonance frequency of the beam, yielding an estimated mass resolution of about $10^{-18}$\,g.

\section{Results}
\begin{figure}
\includegraphics{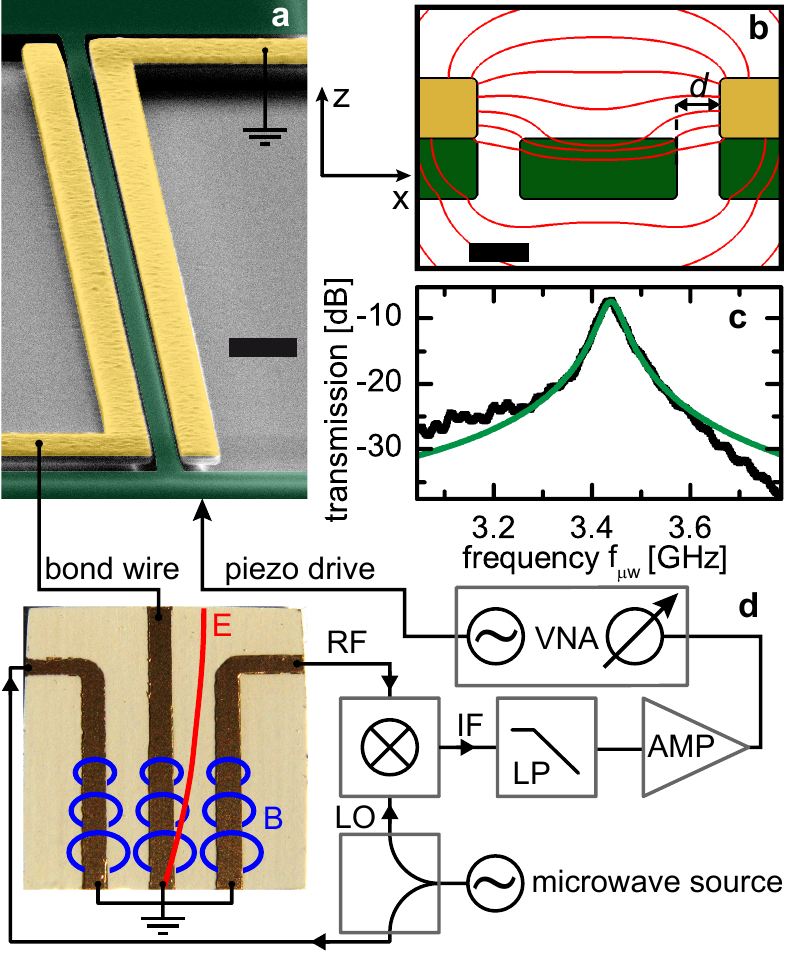}
\caption{\label{fig1}\bf Sample and setup: \sf
SEM micrograph of the 55\,\textmu m long silicon nitride beam (green) flanked by two gold electrodes (yellow) is shown in {\bf a} (scalebar corresponds to 1\,\textmu m).
The schematic cross section of beam and electrodes in {\bf b} (scalebar corresponds to 100\, nm) exhibits a symmetric gap of $d=60$\,nm and includes simulated electric field lines.
The beam is placed just below the electrodes, where its movement in z direction induces the largest modulation of the capacitance.
The electrodes are connected to an electrical $\lambda/4$ microwave cavity via bond wires.
Its transmission spectrum with a Lorentzian fit is shown in {\bf c}, a photo ($8\times 8\,\rm mm^2$) with magnetic field lines (blue) indicating the inductive coupling between the two side electrodes and the central resonator and the electric field distribution (red) in the resonator is included in {\bf d}.
The cavity is pumped by a microwave source, the RF transmission signal is mixed (see methods for the detailed circuit) with a reference signal (LO) such that the mechanical sidebands (IF) are demodulated.
A lowpass filter (LP) is used to remove higher frequency components and the amplified (AMP) sideband signal is fed to a vector network analyser (VNA) which can also drive a piezo to actuate the beam.}
\end{figure}

\textbf{Device and measurement setup:}
Arrays of mechanical beam resonators of different length are fabricated out of a 100\,nm thick pre-stressed silicon nitride film deposited on a fused silica wafer (see \fig{fig1}a for one element as well as methods  and Supplementary Fig. S1).
Each nanomechanical resonator is embedded in a capacitive structure which is part of the resonant LC circuit as sketched in \fig{fig1}a,b and c,d, respectively.
One of the electrodes is connected to an external $\lambda/4$ microstrip cavity (see \fig{fig1}d and methods) with a resonance frequency of $f_c=3.44$\,GHz and a quality factor of 70, the transmission of which is shown in \fig{fig1}c.
In contrast to the more common SiN films on silicon substrates, SiN on fused silica avoids room temperature dissipation of microwave signals by mobile charge carriers and generates an even higher tensile stress in the SiN film\,\cite{Retajczyk1980241}.
Measuring the resonance frequencies of several harmonic modes and fitting these with a simple theoretical model\,\cite{PhysRevLett.105.027205} yields a beam stress of $1.46\pm0.03$\,GPa.
Recently, it has been demonstrated that the tensile stress in a nanomechanical resonator enhances its eigenfrequency and quality factor\,\cite{PhysRevLett.105.027205,verbridge:124304}.
Thus, the observed quality factors are higher than the ones measured with resonators of the same eigenfrequency on a silicon substrate with a prestress of 0.83\,GPa, and beams of the same length have higher resonance frequencies.
Whereas the described scheme has been employed on a range of microwave cavities and nanomechanical resonator arrays, all measurements shown here have been performed on one 55\,\textmu m long beam with a fundamental mechanical resonance frequency of $f_{\rm m}=6.6\,\rm MHz$ and a room temperature quality factor $Q_m=290,000$.

\begin{figure*}[htb]
\includegraphics{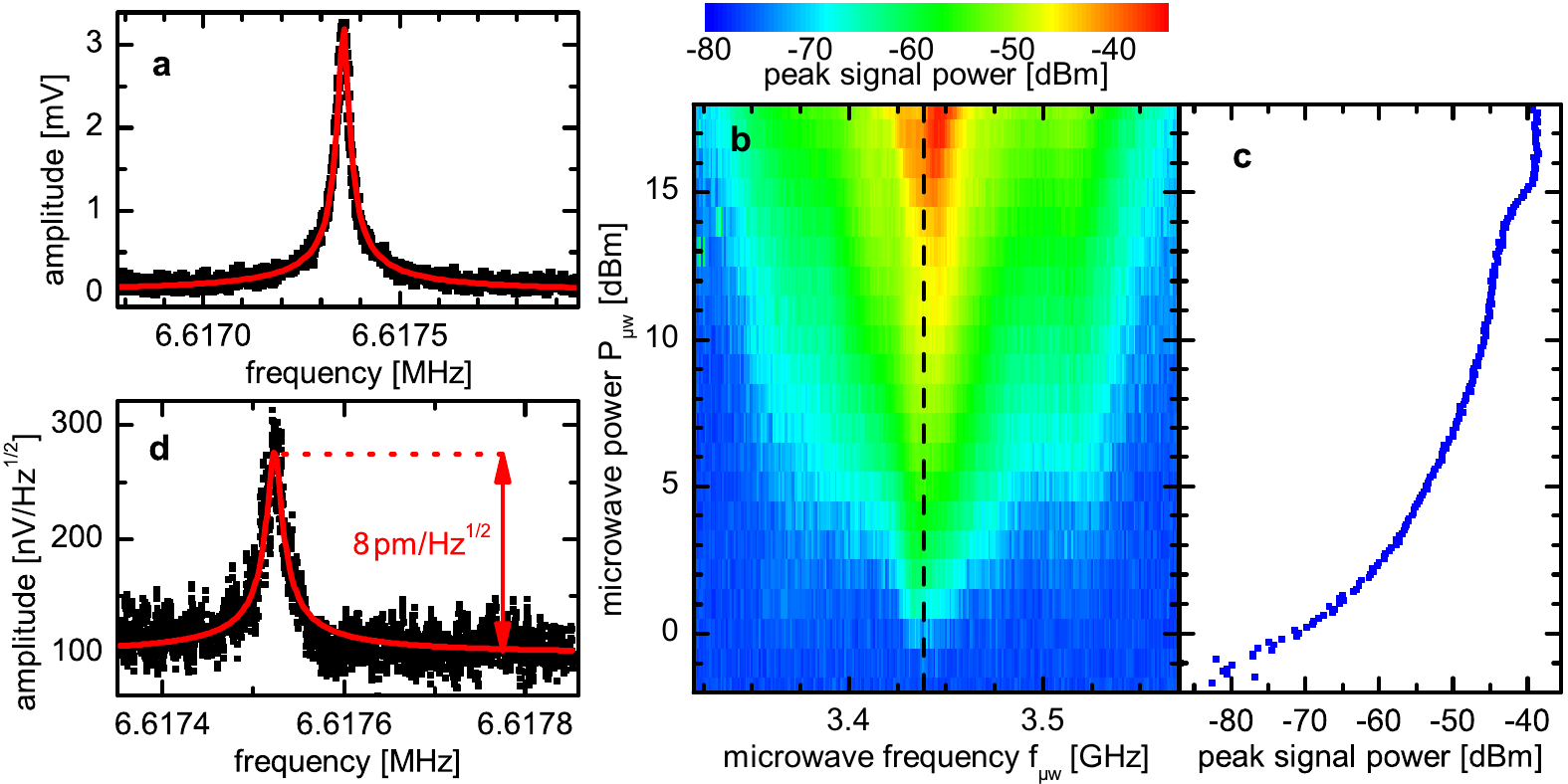}
\caption{\label{fig2}\bf Detection: \sf The driven response of the 55\,\textmu m long beam vibrating with an amplitude of about $6$\,nm at a piezo driving power of -70\,dBm measured with a microwave power of $P_{\mu w}=$18\,dBm at $f_{\mu w}=3.44$\,GHz is shown in {\bf a}.
To  determine the operating regime of the detection scheme as a function of the cavity parameters, microwave frequency and power are systematically varied and the mechanical peak signal power reflecting the squared resonance amplitude (i.e. the maximum of the resonance in {\bf a}) is plotted in {\bf b} for each point. A cut along the microwave cavity resonance (dashed line in {\bf b}) shown in {\bf c} depicts that the detected signal is maximised at cavity powers between 15 and 18\,dBm.
The Brownian motion of the resonator in {\bf d} is employed to deduce the sensitivity of the detection scheme from the noise floor. 
}
\end{figure*}
 
By coupling the mechanical resonator to the microwave cavity, the electrical resonance frequency $f_c$ is periodically modulated, causing sidebands at $f_{\rm c}\pm f_{\rm m}$ in the microwave transmission signal.
These are demodulated, filtered and amplified (see \fig{fig1}d and methods), then fed directly into a vector network analyser, the output of which can be used to excite the mechanical resonator via a piezo inertial drive\,\cite{verbridge:124304}.
The resonator chip glued onto the piezo transducer as well as the cavity are operated in a vacuum chamber at pressures below $5\cdot10^{-4}$ mbar at room temperature.

\textbf{Detection:}
The amplitude of the piezo-driven mechanical resonator (\fig{fig2}a) is probed by monitoring the sideband signal via the demodulated microwave transmission signal.
Please note that all amplitudes in this work are given as half-peak-to-peak values.
A Lorentzian fit is used to extract the mechanical resonance amplitude, which depends on the output power $P_{\mu w}$ and frequency $f_{\mu w}$ of the microwave source.
Figure 2b shows how the operating range of the detection mechanism can be mapped out by systematically varying the microwave parameters.
The plot displays the colour-coded peak signal power of the mechanical spectrum, plotted for every set of $P_{\mu w}$ and $f_{\mu w}$.
Maximum sensitivity is achieved on resonance with the microwave cavity (at $f_{\mu w}=f_c=3.44$\,GHz, dashed line), where the cavity field and its sensitivity to frequency changes are maximised.
The peak signal power in \fig{fig2}c is directly proportional to the microwave power at low levels, but nonlinear effects in the cavity cause the detection efficiency to level off above $P_{\mu w}$=15\,dBm.
Using the optimal operating point of $f_{\mu w}=f_c$ and $P_{\mu w}=18$\,dBm, the thermally induced Brownian motion of the resonator can be easily resolved, as depicted in \fig{fig2}d.
By calculating the thermal amplitude of the beam to be $8\,\rm\frac{pm}{\sqrt{Hz}}$ at room temperature, the observed noise level corresponds to a sensitivity of $4.4\,\rm\frac{pm}{\sqrt{Hz}}$ (see also Supplementary Note 1).
Thus, the full dynamic range of the resonator is accessible since the detection scheme allows to  characterise the resonator response from the thermal motion until the onset of nonlinear behaviour.

As the detected signal is only proportional to the change in capacitance $\frac{dC}{dz}$ caused by a displacement $dz$ and other geometrical parameters, the displacement sensitivity is independent of the mechanical frequency.
However, higher frequency beams imply a reduced electrode length and thus weaker coupling for constant cross section of the detection capacitor.
 The same applies to higher harmonic modes, where only one antinode of odd harmonic modes generates a signal, as the other antinodes cancel each other.
This results in a $\frac{1}{f}$ scaling of the sensitivity in the case of a stressed string, as observed for beams with frequencies between 6 and 60\,MHz (not shown).

 Furthermore, even the in-plane motion of the beam can be detected (not shown).
Considering the electrode geometry shown in \fig{fig1}b, this seems to be suprising at first.
Ideally, the capacitance gradient $\frac{dC}{dx}$ is a parabola such that both a displacement of the beam in positive and negative x direction increases the capacitance symmetrically.
Thus, there should be no signal on the resonance frequency of the mode.
But even small imperfections during sample fabrication lead to a slightly off-center position of the beam and thereby a nonzero capacitance gradient in the x direction such that in-plane modes are accessible, albeit with a lower sensitivity

\begin{figure}
\includegraphics{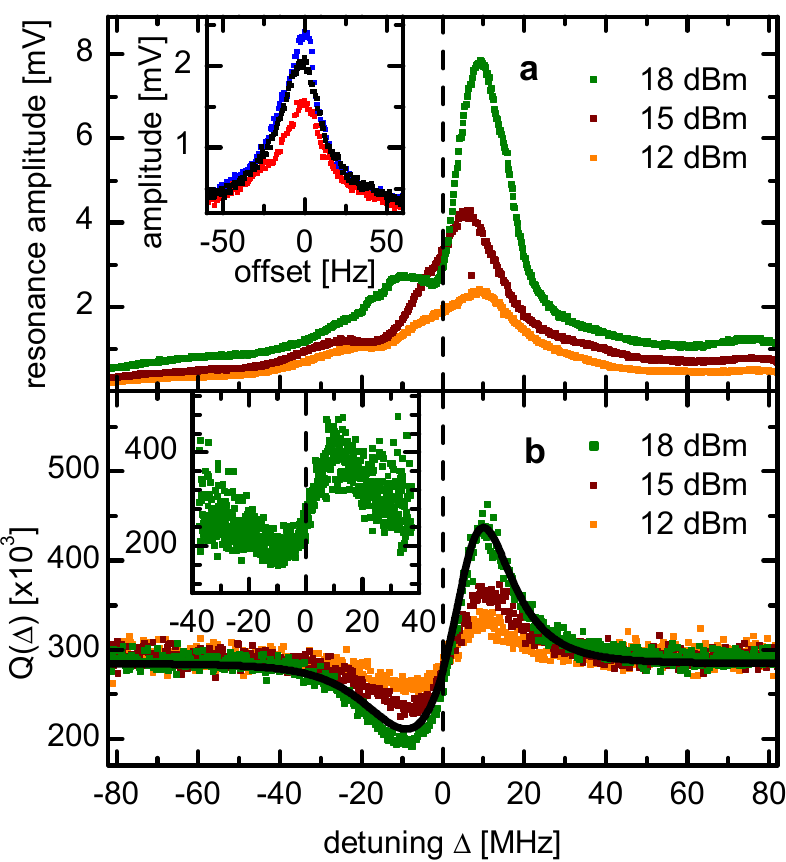}
\caption{\label{fig3}\bf Cavity electromechanics: \sf Resonance amplitude and quality factor of the weakly driven (piezo power of -70\,dBm) mechanical  mode are increased or decreased depending on the microwave detuning $\Delta = f_{\mu w}-f_c$ due to the backaction of the microwave field on the resonator. This is shown in the inset of {\bf a}, comparing resonance curves of the same resonance at a microwave power of 12\,dBm for red, blue or no (red, blue and black points, respectively) detuning.
For clarity, the resonance curves have been plotted versus the frequency offset to the respective (detuning-dependent) resonance frequency.
The detuning dependence of the resonance amplitude (i. e. maximum in inset) is shown for different microwave powers in {\bf a}.
The detuning dependence of the quality factor is depicted in {\bf b} for the same power values, along with a fit to the theoretical model (see Supplementary Note 2). The inset of {\bf b} shows the quality factor of the Brownian motion vs. detuning at 18\,dBm microwave power, the only difference to the main plot is the better signal to noise ratio in the weakly driven case.}
\end{figure}

\textbf{Backaction effects:} For a detuned microwave cavity, the coupling between the cavity and the mechanical resonator gives rise to "opto"\-me\-cha\-ni\-cal effects such as backaction cooling and pumping of the mechanical mode\,\cite{Teufel2011,Chan2011,Metzger2004,Kleckner2006,Schliesser2008,Eichenfield2009a,PhysRevLett.99.137205,PhysRevLett.99.093902,1367-2630-10-9-095002,bib:Gigan2006,bib:Arcizet2006a,hao:113501,Rocheleau2010}.
The signature of these cavity electromechanical effects can already be discerned in the top center portion of \fig{fig2}b and is shown more clearly in \fig{fig3}.
Comparison of the different mechanical resonance curves obtained for negative, positive and no detuning (inset of \fig{fig3}a) shows that both the resonance amplitudes (\fig{fig3}a) and the measured, effective $Q(\Delta)$ (\fig{fig3}b, see Supplementary Note 2) change with detuning.
If the detuning $\Delta=f_{\mu w}-f_c$ between microwave drive and cavity resonance frequency is negative (red detuned), the electrical force produced by the cavity field counteracts the vibrational motion, thereby decreasing its amplitude.
For positive (blue) detuning, the amplitude is increased.
Since the resonance amplitude  depicted in \fig{fig3}a is superimposed with the detuning-dependent sensitivity curve discussed in \fig{fig2}b and therefore distorted, we rather use the  detuning dependence of the quality factor to analyse the data.

The effective $Q(\Delta)$ in \fig{fig3}b clearly shows the expected behaviour: At negative detuning, the additional cavity-induced damping $\Gamma(\Delta)$ is positive, such that the effective damping exceeds the intrinsic value and $Q(\Delta)$ decreases, whereas at positive detuning the opposite occurs, with an optimal detuning of $|\Delta_{opt}|=9$\,MHz.
Fitting the theoretical model (\cite{PhysRevLett.99.093902,1367-2630-10-9-095002,RevModPhys.82.1155} and Supplementary Note 2) to the data measured at several cavity drive powers allows to extract the average coupling factor $g=\frac{\partial f_c}{\partial z}=75\pm5\,\rm\frac{Hz}{nm}$.
The backaction effect is independent of piezo-driven beam actuation as only the effective damping is changed. This is confirmed by repeating the experiment without piezo actuation (inset of \fig{fig3}b). 
A comparison between the weakly driven situation depicted in \fig{fig3}b and the Brownian motion in the inset only shows a significant increase  of the noise in the latter case.
Therefore, all measurements in Figs.\,3\,\&\,4a (except the inset in \fig{fig3}b) were done with a weak piezo actuation of -70\,dBm to operate with an improved signal to noise ratio.

\begin{figure}
\includegraphics{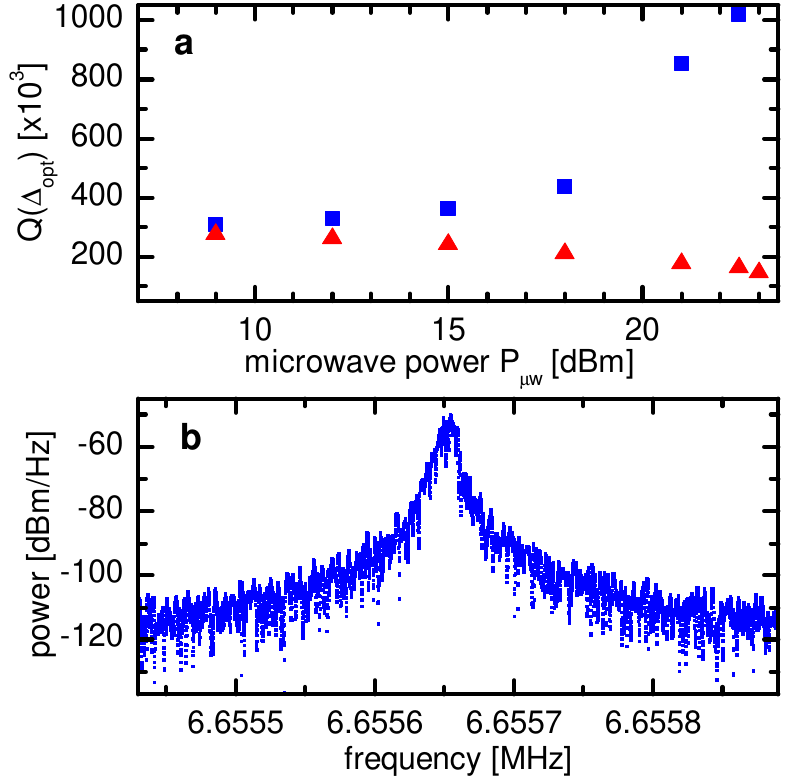}
\caption{\label{fig4}\bf Cavity-induced damping and self-oscillation: \sf Using a blue (red) detuned cavity drive, the amplitude of the beam can be amplified (damped). This effect is controlled via the microwave power $P_{\mu w}$, as shown in {\bf a} (blue squares: blue detuning, red triangles: red detuning) for the optimal $\Delta_{opt}$ of $\pm 9$\,MHz.
By increasing the microwave power to 23\,dBm, the backaction gain caused by the blue-detuned cavity exceeds the intrinsic damping and self-oscillation occurs.
The respective power spectrum in {\bf b} shows a linewidth reduced to 5\,Hz.}
\end{figure}

Increasing the microwave power to 23 dBm, the quality factor can be decreased to half its  initial value with a negative $\Delta$ as shown in \fig{fig4}a, corresponding to an effective mode temperature of 150\,K\,\cite{Metzger2004}.
For positive $\Delta$, the quality factor diverges.
This reflects cavity-driven self-oscillation of the beam,  once the intrinsic damping is canceled by the cavity backaction.
The power spectrum of this oscillation is shown in \fig{fig4}b.
Its linewidth of 5\,Hz, corresponding to an "effective quality factor"\,\cite{FengX.L.:342} of 1.3 million, is limited by the stability of the oscillation frequency, which is mainly affected by fluctuations of the cavity drive.

This ultra-low linewidth is ideally suited for mass sensing applications, giving rise to a mass resolution of about $10^{-18}$\,g (see Supplementary Note 3).
In contrast to single carbon nanotubes, which have been employed to probe masses of $10^{-{22}}$\,g\,\cite{JensenK.:533,doi:10.1021/nl801982v}, the presented scheme can readily be scaled up to a large-scale fabrication process involving many beams. Further improvements can be expected by increasing the electromechanical coupling constant\,\cite{Teufel2011,Regal2008}. This can be achieved by a reduced gap size in the detection capacitor. Beam-electrode-separations of 20\,nm have already been demonstrated\,\cite{doi:10.1021/nl102771p}, which should yield a 10-fold increase in coupling.

Besides the backaction effects, there is a quasistatic electric force acting on the resonator\,\cite{Unterreithmeier2009}.
The electric microwave field between the electrodes polarises the dielectric beam, creating dipoles which are attracted to high electric fields.
This leads to an additional effective spring constant that scales with the square of the field (i. e. with $P_{\mu w}$) and leads to an increased (decreased) restoring force for the out-of-plane (in-plane) mode.
The resulting difference in resonance frequency is clearly visible comparing \fig{fig1}a ($P_{\mu w}$=18\,dBm) to \fig{fig4}b ($P_{\mu w}$=23\,dBm) and can be employed to tune the mechanical eigenfrequency.

\section{Discussion}

 There are only few existing nanomechanical transduction schemes at room temperature providing good integration and scalability to large resonator arrays coupled to a single readout cavity: Photonic circuits\,\cite{Pernice09,PhysRevLett.103.103601,Eichenfield2009} offer  extremely large displacement sensitivities, but are limited by the precise alignment of external components and thus sensitive to vibrations.
On the other hand, adjustment-free schemes such as piezoelectric transduction\,\cite{mahboob:253109} or capacitive detection\,\cite{Knobel2003,Regal2008}, which, in addition, frequently require cryogenics, impose material constraints and can cause additional dissipation\,\cite{Sekaric2002,2011arXiv1111.1703Y}.
In contrast, the  presented dielectric coupling of the nanomechanical resonator to the microwave cavity allows to maintain a large quality factor over a wide temperature range (tested between 4 and 300\,K).
Accordingly, the reported room temperature $Q_m$ of 290,000 of the prestressed SiN-on-fused-silica nanoresonator is, to our knowledge, the highest ever obtained in this frequency range.

In conclusion, we present a room-temperature platform for the sensitive readout, actuation and tuning of nanomechanical resonators.
We  achieve a sensitivity well below the Brownian motion for the fully integrable and robust heterodyne readout of a nanomechanical resonator via a weakly coupled microwave cavity ($g=\frac{\partial f_c}{\partial z}=75\pm5\,\rm\frac{Hz}{nm}$).
 This coupling constant is significantly smaller than the one obtained with capacitively coupled beams\,{\cite{Regal2008} but requires neither cryogenics nor beam metallisation.
The relative cavity frequency shift $\frac{g}{f_c}$ is comparable to typical optical experiments\,\cite{Anetsberger2009,bib:Thompson2008a}, as not only the coupling constant $g$ but also the cavity resonance frequency $f_c$ are orders of magnitude smaller  in the microwave regime.

A major advantage of the presented scheme is the parallel readout of many beams and many modes (higher harmonics as well as in and out of plane) using only a single microwave setup.
Additionally, the cavity backaction can be used to control the amplitude of the resonator, thus allowing to omit the piezo actuator. 
By entering the regime of cavity-pumped self-oscillation a strong and narrow-band signal is generated, perfectly suited for sensing applications requiring a simple resonance frequency readout.
As both resonator and cavity are fabricated reproducibly using standard lithographic processes, inexpensive plug and play NEMS sensor modules using only two microwave connectors to interface them with control electronics can be developed.

\section{Methods}
\textbf{Microwave Setup:}
The microwave cavity is fabricated on a ceramic substrate suitable for high frequency applications (Rogers TMM10) cut to small chips.
Standard optical lithography and wet etch processes are employed to pattern the 17\,\textmu m thick top copper layer, onto which a 150\,nm gold coating is evaporated to avoid corrosion.

The design of the microwave cavity shown in \fig{fig1}d consists of an 8\,mm long and $0.64$\,mm wide center strip which forms the actual $\lambda/4$ resonator.
One end of the strip is grounded, while its other end is connected to the silica chip carrying the mechanical resonators.
Two adjacent strips near the grounded end are used to inductively couple the cavity to the feed lines and measure the transmission signal.
We chose inductive coupling to separate the interface to the chip at the open end from the interface to the feed lines at the grounded end of the $\lambda/4$ resonator.
The length of the two feed lines (6\,mm) and the distance between the striplines (1.2\,mm) were optimised using high-frequency circuit simulations with APLAC for a tradeoff between transmission and quality factor of the cavity.

\begin{figure}[t]
\includegraphics{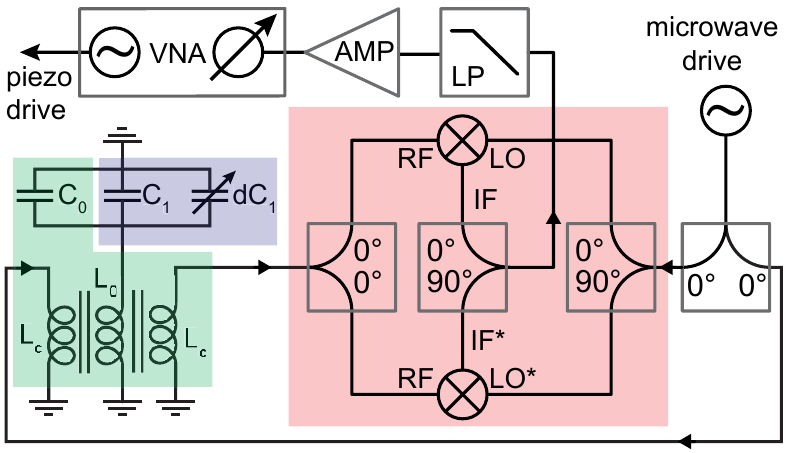}
\caption{\label{fig5}\bf Detailed electrical schematic: \sf Detailed version of Fig.~1d showing an electrical equivalent circuit of the microwave cavity (green box) coupled to the resonator chip (blue box) and the complete IQ mixing circuit (red box).}
\end{figure}

 Figure~5 shows a detailed version of the simplified electrical circuit depicted in \fig{fig1}d.
On the left, the electrical equivalent circuit of the $\lambda/4$ microstrip cavity including both feed lines and the coupling to the chip is depicted.
The actual cavity consists of the inductance $\rm L_0$ of the 8\,mm long copper strip and its capacitance $\rm C_0$ to the ground plane on the bottom of the circuit board.
The inductive coupling between $\rm L_0$ and the two feed lines with inductances $\rm L_c$ provides the external interface to the cavity.
The resonator chip is connected to the open end of the cavity, it adds a static capacitance $\rm C_1$ and a time-dependent contribution $\rm dC_1(t)$, which oscillates with the actual beam displacement.
A very rough estimate for $\rm L_0$ and $\rm C_0$ can be obtained from microstrip theory, giving values of 1\,pF and 3\,\textmu H (this would result in $f_c=\frac{1}{2\pi\sqrt{\rm L_0C_0}}=2.9$\,GHz), neglecting the effects of the microstrip ends.
As the resonance frequency of the bare microwave cavity is about 5\% higher than with the bond wire and chip connected, $\rm C_1$ must be less than 10\% of $\rm C_0$, as the bond wire also adds some inductance to the circuit (not shown in the figure).

 A microwave tone is applied to one port of the cavity, thus creating a phase-modulated signal at the other port of the cavity caused by the alternating capacitance $\rm dC_1$.
As the transmission through the cavity also adds some detuning-dependent phase to the microwave tone, directly mixing the cavity output with the drive tone would result in a detuning-dependent phase difference between the two signals.
This difference would need to be compensated for by an additional phase shifter, which had to be adjusted for maximum signal at every drive frequency.
To avoid this tedious procedure, we use an IQ mixer shown in the red box.
It consists of a 0°/0° and a 0°/90° power splitter as well as two mixers.
The reference signal coming from the microwave generator is split into two parts with a 90° phase shift to each other.
By mixing these two signals (LO and LO*) with the RF transmission signal of the cavity, two intermediate freqency signals (IF and IF*) are created.
Depending on the phase of the two input signals of the IQ mixer, at least one of these signals is always non-zero, and their (phase-correct) sum is completely independent of the phase relation of the input signals.
Thus, by combining the two demodulated quadrature components with another 0°/90° power splitter and blocking the higher frequency mixing products, amplitude and phase of the mechanical signal are reconstructed.

The noise background of this signal is primarily caused by the phase noise of the frequency generator driving the electrical cavity, causing more background noise with increasing power.
Therefore we use a Rohde\&Schwarz SMA100A signal generator with extremely low phase noise below $-150$\,dBc at 10\,MHz offset.
In order to preserve the low noise level, the demodulated sidebands are amplified with a 35\,dB preamplifier with a noise figure of 1.3\,dB.
The output of this amplifier is either directly connected to a spectrum analyser (to quantify the Brownian motion) or amplified by another 30\,dB and fed to a network analyser (in case of the driven measurements).

\textbf{Resonator fabrication:}
The samples are fabricated on 500\,\textmu m thick fused silica wafers which are coated with a 100\,nm thick commercial high quality LPCVD layer of pre-stressed silicon nitride.
$5\times 5\rm\, mm^2$ large chips are cut from the wafer.
To enable electron beam lithography on these non-conductive substrates, 2\,nm of chromium are evaporated onto the PMMA resist before exposure and removed prior to developing.
E-beam lithography and standard lift-off processes are used to define the gold electrodes and a thin cobalt etch mask protecting the beam.
The subsequent ICP reactive ion etch using $\rm SF_6$ and $\rm Ar$ removes the silicon nitride which is not protected by a metal layer.
The final hydrofluoric acid wet etch removes the cobalt and releases the beams, while the gold electrodes use chromium as an adhesion layer and are not attacked by the acid.
Finally, the chips are blow-dried with nitrogen, glued to the piezo and a wire-bonder is used to connect them to the microwave cavity.
 All these processing steps use industry-standard techniques, so a large-scale fabrication of inexpensive sensor modules should be within reach.

Each mechanical resonator chip contains multiple beams with their respective electrodes, all shunted between two bond pads which are used to connect the chip to the microwave cavity.
One design with big variations in the beam length is shown in Supplementary Fig. S1.
It is also possible to use designs with very small length differences in the order of 100\,nm, allowing to address many mechanical resonances by frequency division multiplexing in a narrow frequency band.

 \textbf{Resonator chip design:}
In order to choose the sample design with the highest coupling between the electrical cavity and the mechanical resonator, several simulations of the electrode configuration using COMSOL Multiphysics were conducted.
The electrodes were patterned directly onto the SiN film to induce a maximal capacitance variation  with beam displacement.
We decided to put the gold electrodes on top of the silicon nitride layer and thereby deposit them before the reactive ion etch step, in contrast to our earlier designs where the gold was evaporated onto the remaining silicon dioxide covering the silicon substrate below the resonator\,\cite{Unterreithmeier2009}.
In these previous designs, the vertical separation between the beam and the silicon dioxide layer had to exceed 250\,nm to achieve sufficient underetching of the beam.
Thus the new design allows for much smaller overall beam-electrode separations, resulting in a larger effect of the beam motion on the capacitance.
Further simulations with this principal geometry varied several other parameters.
These primarily show an $\frac{1}{d}$ scaling between the lateral beam-electrode distance $d$ (see Fig.~1b) and the capacitance change per nanometer beam displacement, as expected for a capacitive interaction.

As the beams tend to stick to the side electrodes at very low gap sizes, the fabrication of smaller defect-free gaps by conventional SEM lithography and dry\,\&\,wet-etching was not successful.
We have investigated devices with gap widths varying between 110\,nm and \,60nm. Supplementary Figure S2 shows the coupling strength (black squares) extracted from the quality factor versus microwave frequency curves, as shown in Fig.~3b, for different values of $d$.
The dotted red curve depicts the capacitance gradients obtained from the simulation multiplied with a scaling factor to fit the measured coupling strength.
The dependence of the coupling strength on the gap size in the measurements is qualitatively reproduced by the simulations.

\section{Competing financial interests}

The authors declare no competing financial interests.

\section{Acknowledgments}
Financial support by the Deutsche Forschungsgemeinschaft via Project No. Ko 416/18, the German Excellence Initiative via the Nanosystems Initiative Munich (NIM) and LMUexcellent, the German-Israeli
Foundation (G.I.F.), as well as the European Commission under the FET-Open project QNEMS (233992) is gratefully acknowledged.
We would like to thank Florian Marquardt and Johannes Rieger for stimulating discussions.


%

\onecolumngrid
\clearpage
\renewcommand{\figurename}{Supplementary Figure}
\renewcommand{\thefigure}{S\arabic{figure}}
 \renewcommand{\theequation}{S\arabic{equation}}
 \setcounter{figure}{0}
\setcounter{equation}{0}
\renewcommand{\fig}[1]{Supplementary Figure~\ref{#1}}
\renewcommand{\eq}[1]{equation~(\ref{#1})}

\section{Supplementary information}

\begin{figure}[ht]
\includegraphics{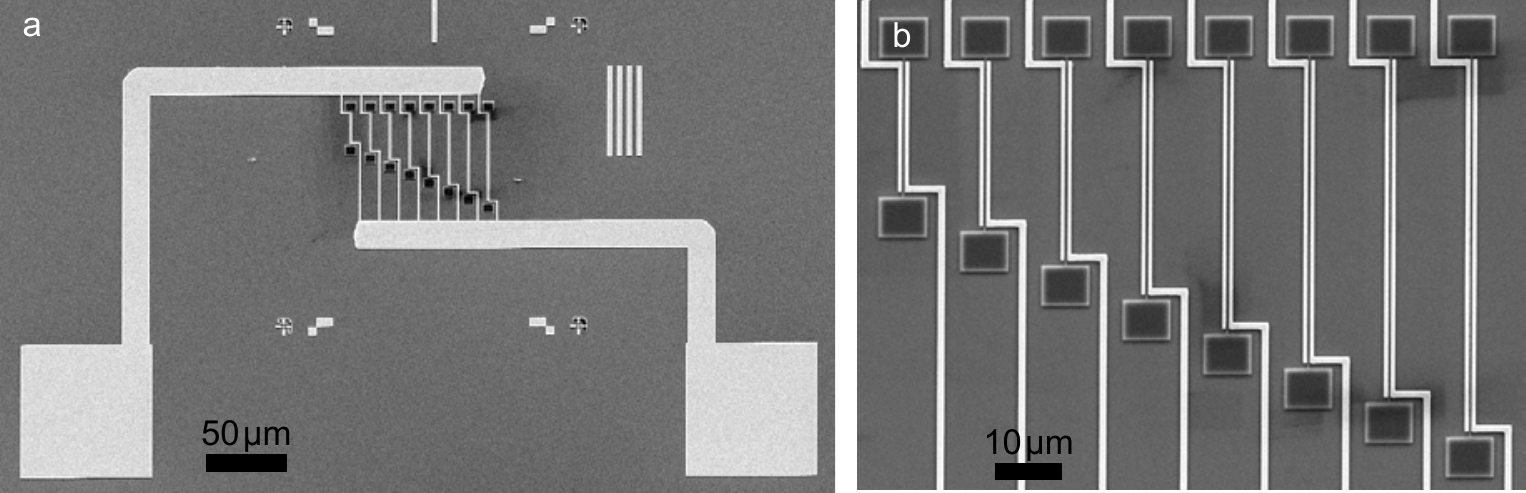}
\caption{\label{s1}\textbf{Mechanical resonator array:} A SEM micrograph of the entire mechanical resonator structure including the bond pads is shown in {\bf a}. The central array of eight resonators with lengths ranging from 20 to 55\,\textmu m is shown in more detail in {\bf b}.}
\end{figure}

\newpage

\begin{figure}
\includegraphics{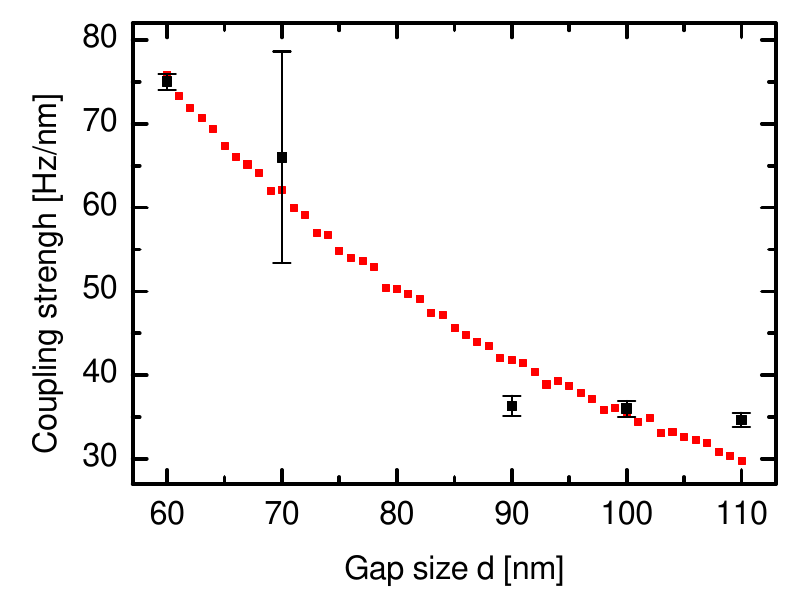}
\caption{\label{s2}\textbf{Coupling strength versus gap distance:}
The black markers show the measured values of $g$ for several different beams, including their error bars.
The large error at a gap size of 70\,nm is caused by the low quality factor of this beam, presumably contaminated by etch residuals.
The red dots are the capacitance gradients obtained from the simulation, multiplied by a scaling factor.
There is a qualitative agreement between the simulation results and the measured data.}
\end{figure}

\clearpage

\subsection*{Supplementary Note 1: Displacement sensitivity}

The displacement sensitivity is calibrated by calculating the amplitude distribution of the thermally induced Brownian motion of the beam\,[37].
The average kinetic energy of a harmonic oscillator in thermal equilibrium is
$$\langle E_{kin}\rangle=\frac{1}{2}k_BT=\frac{1}{4}m_{eff}\int\limits_0^\infty A^2(f)f^2df
$$
where $A(f)$ designates the frequency-dependent displacement of the resonator, $k_B$ is the Boltzmann constant, $T$ the ambient temperature and 
$$
m_{eff}=\frac{m}{2}=\frac{\rho\cdot l\cdot w\cdot t}{2}=1.9\cdot10^{-12}\,\rm g
$$
the effective mass of a string with a sinusoidal oscillation profile, using the densitiy of silicon nitride $\rho=2,600\,\rm kg/m^3$, the beam length $l=55$\,\textmu m, width $w=260$\,nm and thickness $t=100$\,nm.
By assuming a Lorentzian distribution with the measured beam center frequency $f_m$ and quality factor $Q_m$ for $A(f)$, the amplitude of the spectral displacement distribution $A(f)$ can be calculated, yielding a peak value of $8\,\rm\,\frac{pm}{\sqrt{Hz}}$.

By definition, the mechanical amplitude corresponding to a signal-to-noise ratio of unity defines the displacement sensitivity of the detection scheme.
The peak amplitude of the fit in Fig.~2d of the main text is $276\,$nV, the noise level $98\,$nV, yielding a SNR of 1.8.
We thus achieve a sensitivity of $\frac{8\rm\,\frac{pm}{\sqrt{Hz}}}{1.8}=4.4\rm\,\frac{pm}{\sqrt{Hz}}$.

\subsection*{Supplementary Note 2: Theory of cavity-induced damping}

A full quantum theory of cavity-assisted sideband cooling is given by Marquardt et al.\,[27].
In the following, his results are converted such that only experimentally accessible parameters enter, similar to the representation in the work of Teufel et al.\,[28].
The additional mechanical backaction damping exerted by the cavity is given as\,[27]
\begin{equation}
\label{gamma}
\Gamma=\frac{1}{\hbar^2}\left[S_{FF}(\omega_m)-S_{FF}(-\omega_m)\right]x_{ZPF}^2
\end{equation}
with $\omega_m=2\pi f_m$ and the zero point fluctuation $x_{ZPF}=\sqrt{\frac{\hbar}{2m_{eff}\omega_m}}$.
Furthermore, the radiation pressure power spectrum is 
\begin{equation}
\label{sff}
S_{FF}(\omega)=\hbar^2A^2\bar{n}\frac{\kappa}{(\omega+\Delta')^2+(\kappa/2)^2}
\end{equation}
using the average number of cavity photons $\bar{n}$, the cavity damping $\kappa=\frac{2\pi f_c}{Q_c}=\frac{\omega_c}{Qc}$, the coupling constant $A=\frac{\partial\omega_c}{\partial z}$ and the angular frequency detuning $\Delta'=2\pi(f_{\mu w}-f_c)$.

Equation~\ref{gamma} can then be expressed as
\begin{equation}
\Gamma=\frac{2\hbar A^2\bar{n}}{m_{eff}\omega_m}\left[\frac{\kappa}{\kappa^2+4(\Delta'+\omega_m)^2}-\frac{\kappa}{\kappa^2+4(\Delta'-\omega_m)^2}\right].
\end{equation} 
The power lost in the cavity is $\kappa E_{\rm stored}=\kappa \bar{n}\hbar\omega_c$. In a steady state, this has to equal the power coupled into the cavity via the feed lines.
Assuming that the cavity losses $\kappa$ are only due to energy loss into the two symmetric feed lines (and thereby the coupling constant between one feed line and the cavity is $\kappa/2$), equation E47 in the supplement of Clerk et al.\,[29] gives the relation
\begin{equation}
\label{pinc}
P_{inc}=\hbar\omega_c\bar{n}\frac{\kappa}{2} \Rightarrow \bar{n}=\frac{2P_{inc}}{\hbar\omega_c\kappa}
\end{equation}
for a two-sided cavity where the power is provided only via one of the two feed lines. Furthermore, this has to be multiplied with the cavity resonance $\frac{\kappa^2}{\kappa^2+4\Delta'^2}$.
Thus, the electromechanically induced damping can be written as
\begin{equation}
\label{gammaomega}
\Gamma(\Delta')=\frac{4P_{inc}A^2\kappa^2}{m_{eff}\omega_m\omega_c(\kappa^2+4\Delta'^2)}\left[\frac{1}{\kappa^2+4(\Delta'+\omega_m)^2}-\frac{1}{\kappa^2+4(\Delta'-\omega_m)^2}\right]
\end{equation}
or converted in frequency units
\begin{equation}
\label{gammaf}
\Gamma(\Delta)=\frac{4P_{inc}g^2\kappa^2}{m_{eff}f_mf_c(\kappa^2+16\pi^2\Delta^2)}\left[\frac{1}{\kappa^2+16\pi^2(\Delta+f_m)^2}-\frac{1}{\kappa^2+16\pi^2(\Delta-f_m)^2}\right]
\end{equation}
using the normal frequency detuning $\Delta=f_{\mu w}-f_c$ and the coupling constant $g=\frac{\partial f_c}{\partial z}$.
The angular frequency result of \eq{gammaomega} agrees with the formula used in\,[28] to within a factor of two.
This is a consequence of the one-sided cavity and therefore a twice as effective input power coupling employed in the work of Teufel et al., leading to an additional factor two in the relation between input power and the number of cavity photons\,(\ref{pinc} and\,[29]).

Using \eq{gammaf}, the effective quality factor $Q$ of the mechanical mode depending on both the intrinsic damping $\gamma_m=\frac{2\pi f_m}{Q_m}$ and on the detuning $\Delta$ of the electrical cavity can be calculated:
\begin{equation} \label{Q}
Q(\Delta)=\frac{2\pi f_m}{\gamma_{total}}=\frac{2\pi f_m}{\gamma_m+\Gamma(\Delta)}.
\end{equation}
This formula was used to extract the coupling constant from the measured data sets of quality factors versus detuning and is shown as a black line in Fig. 3b of the main text.
Note that the incident microwave power is the power in the feedline and not the generator output power, therefore $P_{inc}=0.32\cdot P_{\mu w}$, as 5 dB are lost in the cables and power splitter between the microwave source and the cavity.

\subsection*{Supplementary Note 3: Mass sensitivity}

The mass sensitivity is estimated as follows:
A small mass change $\delta m$ of the effective mass $m_{eff}$ of the resonator will lead to an eigenfrequency shift$$
\frac{\delta f}{f_m}=\frac{\delta m}{m_{eff}}.
$$
We assume that a resonance frequency shift $\delta f=2.5$\,Hz equaling half the self-oscillation linewidth can be resolved.
The effective mass of an oscillating string with a sinusoidal mode profile is
$$m_{eff}=\frac{m}{2}=\frac{\rho\cdot l\cdot w\cdot t}{2}=1.9\cdot10^{-12}\,\rm g$$ using the densitiy of silicon nitride $\rho=2,600\,\rm kg/m^3$, the beam length $l=55$\,\textmu m, width $w=260$\,nm and thickness $t=100$\,nm.
For a resonance frequency $f_m=6.6$\,MHz, the mass sensitivity thus is $\delta m=7\cdot 10^{-19}$\,g in the center of the beam.
By monitoring several harmonics of the beam, it should also be possible to detect the position of the added mass\,[38].

\subsection*{Supplementary References}
\setlength{\parindent}{0pt}

[37] J. L. Hutter and J. Bechhoefer, Review of Scientific Instruments \textbf{64}, 1868 (1993).\newline
[38] S. Dohn, W. Svendsen, A. Boisen, and O. Hansen, Review of Scientific Instruments \textbf{78}, 103303 (2007).

\end{document}